\documentclass[twocolumn,epsfig,aps,prb]{revtex4}

\begin{document}

\title{Photoluminescence from a gold nanotip in an accelerated reference frame.}

\author{Igor I. Smolyaninov}
\affiliation{Department of Electrical and Computer Engineering, University of Maryland, College Park, MD 20742, USA}

\date{\today}

\begin{abstract}
Photoluminescence from a gold nanotip, which is induced by surface plasmons propagating over a curved tapered nanotip surface is considered in a co-moving accelerated reference frame. Similar to the surface-enhanced Raman scattering (SERS) effect, nonlinear optical mixing of the surface plasmons with the Unruh quanta is supposed to be enhanced by many orders of magnitude.
\end{abstract}


\maketitle

In 1974 Hawking showed \cite{1} that black holes evaporate by emission of thermal radiation. At about the same time Unruh \cite{2}, Moore \cite{3}, Fulling \cite{4}, Davies \cite{5}, and DeWitt \cite{6} introduced a closely related family of effects, which became known as the Unruh effect. They demonstrated that for an accelerating observer vacuum should look like a bath of thermal radiation with temperature $T_U$ defined as

\begin{equation}
\label{eq1}
kT_U=\frac{\hbar a}{2\pi c},
\end{equation}
 
where $a$ is the observer acceleration. While well-established theoretically, this effect is very difficult to observe in the experiment. An observer accelerating at $a=g=9.8 m/s^2$ should see vacuum temperature of only $4\times 10^{-20} K$. Over the past years quite a few proposals were introduced on how to observe the Unruh effect and the Hawking radiation. For example, it was suggested by Schwinger \cite{7} and later Eberlein \cite{8} that the closely related dynamical Casimir effect is responsible for sonoluminescence. However, no unambiguous experimental demonstration has been reported yet. In this paper we argue that the recently detected infrared (IR) photoluminescence from a gold nanotip \cite{9}, which is mediated by surface electromagnetic waves (surface plasmon-polaritons \cite{10}) propagating over a curved tapered metal nanotip surface, can be considered as a nonlinear optical effect closely related to the Unruh and Hawking effects. 

The experimental observations in ref.\cite{9} (considered in the laboratory reference frame) can be summarized as follows. When a gold nanotip is illuminated by a short femtosecond laser pulse, surface plasmon polaritons (SPP) are excited on the tip surface. A broad IR photoluminescence which power depends linearly on the power of the excitation light is observed when these plasmons decay. The observed photoluminescence is very sensitive to the geometrical shape of the gold nanotip. Experimental observations may be explained by a series of electron-electron, electron-plasmon, electron-photon, and plasmon-plasmon interactions. The authors of ref.\cite{9} consider a three-step process involving these interactions as a potential explanation, even though they admit that every step of the proposed process is strongly prohibited by one or another symmetry consideration. Note that the term \lq\lq nanotip\rq\rq is used in the literature to describe metal tips which are smaller than a micrometer. Therefore, classical electrodynamics may be used to describe electromagnetic eigenmodes of these objects.

In this paper we argue that it may be interesting to consider the IR photoluminescence in the reference frame, which is co-moving with a surface plasmon wave packet excited on the gold nanotip surface. Let us show qualitatively that the SPPs which live on the metal tip surface represent a very good system for experimental detection of the Unruh-Hawking-like effects. Surface plasmons are collective excitations of the conduction electrons and the electromagnetic field on the surface of such good metals as gold and silver \cite{10}. They propagate along metal surface at a speed, which is below (but comparable) to the speed of light. Since the SPP momentum is larger than the momentum of a free space photon of the same frequency, plasmons are effectively decoupled from our own three-dimensional space (if the metal surface is not too rough). They live in their own curved two dimensional (2D) spaces, which are defined by the shape of the metal surface. Since SPP wavelength $\lambda_p$ may be quite short (it may fall into the sub-100 nm range \cite{10}), a three dimensional (2D+1T) space-time exhibiting very large curvature (with the radius of curvature $R\sim 1\mu m$) may be emulated in experiments with surface plasmon-polaritons. Since $\lambda_p \ll R$ (which corresponds to the ray optics approximation), a SPP wave packet propagating over a curved metal surface with radius $R$ may be considered as a relativistic massive quasi-particle, which lives in a strongly-curved three-dimensional space-time. The Unruh temperature of vacuum perceived by such a particle during its accelerated propagation along the curved metal-vacuum interface of a tapered nanotip may become very high (similar to the Unruh temperature of vacuum perceived by a photon quasi-particle propagating in a tapered waveguide, which was considered in ref. \cite{11}). The characteristic acceleration may be estimated either from the geometry of the taper \cite{11}, or from the typical radius of curvature $R$ of the nanotip using Eq.(1). A typical centripetal acceleration, which is experienced by surface plasmon-polaritons propagating with velocity $\sim c$ over a curved metal surface with a local radius of curvature $R\sim 1\mu m$ may reach values of the order of $a\sim c^2/R\sim 10^{23} m/s^2$ or about $10^{22}g$. This acceleration exceeds the free fall acceleration near a typical solar-mass black hole by 9 orders of magnitude. According to Eq.(1), this range of accelerations corresponds to the Unruh temperature of vacuum $T_U\sim 400K$, which is large enough to manifest itself in optical experiments performed with surface plasmon-polaritons. Moreover, since a SPP wave packet (and any observer co-moving with the same acceleration) perceives vacuum as a thermal bath with temperature $T_U$, every object which is immersed in this thermal bath should have the same temperature as the bath. Otherwise, from the point of view of an accelerated observer there would be a net energy flow to every body immersed in vacuum. Hence, a SPP wave packet perceives the gold nanotip as having the same temperature $T_U\sim 400K$ as vacuum. In addition, from the same condition of thermal equilibrium, we must conclude that from the point of view of the accelerating observer the SPP field on the surface of a gold nanotip should be in the thermal state with temperature $T_U$. As a result, we conclude that the SPP wave packet propagating over the curved metal surface perceives this surface as very hot. 

Let us assume that a gold nanotip is illuminated with a very short (femtosecond) pulse of external laser light (Fig.1) with intensity $I$ and frequency $\omega$. If the surface of metal particle is rough in some given location, illumination of this rough area will cause excitation of a surface plasmon-polariton wave packet \cite{10}, which would start to propagate along the gold nanotip surface. Let us consider nonlinear optical effects associated with this propagation from the point of view of an observer, which has the same instantaneous acceleration as the acceleration of the SPP wave packet. In this reference frame the IR photoluminescence effect will arise from the nonlinear optical mixing of the SPPs excited by the laser with all the thermal Unruh quanta of the electromagnetic field on the surface and around the gold nanotip, such as the thermal surface plasmons and the thermal photons at temperature $T_U$. This nonlinear mixing will be enhanced due to surface roughness, similar to the well-known effect of surface enhanced Raman scattering (SERS), which causes the down-shifted $\omega -\omega_0$ photons to appear (where $\omega_0$ represents the frequency of some characteristic excitation of the illuminated system). In our case $\omega_0=\omega_U$ represents the characteristic frequencies of the thermal photons and plasmons at temperature $T_U$. In the laboratory frame the so generated photons would look like infrared photoluminescence. The power of this photoluminescence would depend linearly on the power of the laser light illuminating the curved metal surface (in accordance with the experimental observations in ref.\cite{9}). Since typical enhancement factors of SERS reach up to fifteen orders of magnitude \cite{12}, this configuration appears extremely favorable for the experimental observation of the Unruh-Hawking-like effects. This experimental geometry takes advantage of very large accelerations $\sim 10^{22}g$ of SPPs and huge enhancement factors ($\sim 10^{15}$) of SERS. Thus, the qualitative picture of the effect in question seems to be clear.

Let us estimate the intensity of IR photoluminescence $I_{PL}$ which may result from this effect. It may be written as

\begin{equation}
\label{eq2}
I_{PL}\sim 4\chi ^{(2)2}I\times I_U,
\end{equation}

where $\chi ^{(2)}$ is the effective second order susceptibility of the rough gold surface, and $I_U$ is the field intensity of the thermal quanta observed in an accelerated reference frame, which may be estimated using the 2D equivalent of the Planck's law of the black body radiation. In order to simplify calculations let us assume that the gold nanotip is shaped as a pseudosphere \cite{13}. Pseudosphere is the simplest surface exhibiting hyperbolic geometry. Its Gaussian curvature is negative and constant everywhere. The pseudosphere is produced as a surface of revolution of the tractrix, which is the curve in which the distance $r$ from the point of contact of a tangent to the point where it cuts its asymptote is constant. The Gaussian curvature of the pseudosphere is $K=-1/r^2$ everywhere on its surface. It is quite apparent from comparison of Figs.2(a) and (b) that the pseudosphere may be used as a good approximation of the shape of a typical metal nanotip. The surface of the pseudosphere may be understood as a sector of the hyperbolic plane \cite{13}, an infinite surface with the metric     

\begin{equation}
\label{eq3}
ds^2=a^2(d\Theta^2+sinh^2\Theta d\Phi^2),
\end{equation}

which (as shown by Hilbert) cannot be entirely fitted isometrically into the Euclidean space. If the edges of the sector are glued together along the geodesics, the pseudospherical shape is obtained. Thus, the small-scale local metric on the surface of the pseudosphere is described by equation (3) everywhere. SPP modes on the surface of a metal nanotip shaped as a pseudosphere have been described in ref.\cite{14}. A SPP dispersion curve on the surface of the pseudosphere is shown in Fig.3. (it is derived from eqs. (12) and (13) of ref.\cite{14}). Note that the nonzero Gaussian curvature of the surface results in the low-frequency energy gap in the spectrum of SPPs, which makes them behave as massive quasi-particles (a similar effect is well-known e.g. in physics of carbon nanotubes). In the lossless approximation the wavevector of plasmons diverge near the frequency of the surface plasmon resonance $\omega_{sp}$. According to ref.\cite{15} the short-wavelength plasmon modes near the resonance provide the main contribution to the SPP density of states near $\omega_{sp}$. On the other hand, in the intermediate frequency range which corresponds to $kT_U$, the SPP dispersion law is close to linear. Thus, the \lq\lq spectral radiance\rq\rq $I(\omega_U, T_U)$ of the SPP field on a smooth metal surface is

\begin{equation}
\label{eq4}
I(\omega _U ,T_U) = \frac{\hbar\omega_U^2}{8\pi^2c}\frac{1}{(e^{\frac{\hbar\omega_U}{kT_U}}-1)}\approx\frac{\hbar\omega_U^2}{8\pi^2c}e^{-\frac{\hbar\omega_U}{kT_U}}
\end{equation}

Under the experimental conditions described in Fig.4 of Ref.\cite{9}(excitation at $\omega =780$ nm, broad photoluminescence peak around $\omega -\omega _U=900$ nm) the exponential term in eq.(4) is of the order of $10^{-2}$ (assuming $T_U\sim 400K$). Thus, relatively high Unruh temperature perceived by the SPPs propagating over a curved metal surface makes the effect described by eq.(2) non-negligible. In addition, similar to the theory of SERS effect, the localized surface plasmons (LSP) may be expected to enhance this effect considerably. The enhancement occurs when the frequency of incident light coincides with the frequencies of LSPs on a rough metal surface \cite{16}. As a result, the excitation field $I$ is locally enhanced by a factor of $E^2$. Furthermore, the $I_U$ field may be locally enhanced by the same mechanism, providing the total enhancement by a factor of $E^4$, similar to SERS. The net result of both enhancement mechanisms may reach as much as $10^{14}-10^{15}$ (see Ref.\cite{12}). These enhancement factors contribute to the typical values of effective $\chi ^{(2)}$ of a rough gold surface. Taking into account the typical values of the effective $\chi ^{(2)}\sim 2\times 10^{-6}$ esu observed in the second harmonic generation (SHG) and SERS experiments performed on rough gold surfaces \cite{17}, the number of photoluminescence quanta described by eq.(2) may reach $2\times 10^4$ IR photons per second at 100 MW/cm$^2$ peak intensity of 780 nm femtosecond pulses used in \cite{9}. This numerical estimate coincides with the typical number of infrared photons observed in the experiments described in Ref.\cite{9}. 

In conclusion, we have provided reasonable arguments demonstrating that the Unruh-Hawking-like nonlinear optical effect may have already been observed in the IR photoluminescence experiments performed with gold nanotips and described in Ref.\cite{9}. Alternatively, the discussed experimental arrangements for detection of the Unruh-Hawking quanta in a nonlinear optical mixing experiment look realistic enough to justify an experiment, which is specially devoted to this purpose.

\begin{figure}
\begin{center}
\end{center}
\caption{The picture of infrared photoluminescence from the point of view of an observer, which has the same instantaneous acceleration as the acceleration of the SPP wave packet. In this reference frame an observer will see surface-enhanced nonlinear optical mixing of the externally excited SPP field with all the thermal Unruh quanta of the electromagnetic field around the metal particle, such as the thermal surface plasmons and the thermal photons. }
\end{figure}

\begin{figure}
\begin{center}
\end{center}
\caption{(a) Electron microscope photo of an etched metal tip of a scanning tunneling microscope. (b) Pseudosphere, which is obtained as a surface of revolution of a tractrix, may be used as a good approximation of the shape of a metal tip of a scanning probe microscope.}
\end{figure}

\begin{figure}
\begin{center}
\end{center}
\caption{A typical dispersion curve of surface plasmons on a curved metal surface. In the lossless approximation the wavevector of plasmons diverge near the frequency of surface plasmon resonance. }
\end{figure}


\begin{references}

\bibitem{1} S.W. Hawking, Nature 248, 30 (1974).

\bibitem{2} W.G. Unruh, Phys.Rev.D 14, 870 (1976).

\bibitem{3} G. Moore, J. Math. Phys. 11, 2679 (1970).

\bibitem{4} S.A. Fulling, Phys. Rev. D 7, 2850 (1973).

\bibitem{5} P.C.W. Davies, J. Phys. A 8, 609 (1975).

\bibitem{6} B.S. DeWitt, Phys. Rep. 19, 295 (1975).

\bibitem{7} J. Schwinger, PNAS 89, 4091 (1992); J. Schwinger, PNAS 89, 11118 (1992).

\bibitem{8} C. Eberlein, Phys. Rev. Lett. 76, 3842 (1996); C. Eberlein, Phys. Rev. A 53, 2772 (1996).

\bibitem{9} M.R. Beversluis, A. Bouhelier, and L. Novotny, Phys.Rev.B 68, 115433 (2003).

\bibitem{10} A.V. Zayats, I.I. Smolyaninov, and A. Maradudin, Physics Reports, 408, 131 (2005).

\bibitem{11} I.I. Smolyaninov, Physics Letters A 372, 5861 (2008).

\bibitem{12} S. Nie and S.R. Emory, Science 275, 1102 (1997).

\bibitem{13} R. Bonola, Non-Euclidian Geometry (Dover, New York, 1955).

\bibitem{14} I.I. Smolyaninov, Q. Balzano, and C.C. Davis, Phys.Rev.B 72, 165412 (2005).

\bibitem{15} I.I. Smolyaninov, Phys.Rev.Lett. 94, 057403 (2005).

\bibitem{16} M. Moskovits, Surface-Enhanced Raman Spectroscopy: a Brief Perspective. In Surface-Enhanced Raman Scattering – Physics and Applications, (Springer, Berlin, 2006) pp 1-18.

\bibitem{17} K. Tsuboi, $et$ $al.$, J. Chem.Phys. 125, 174703 (2006).  

\end{references}
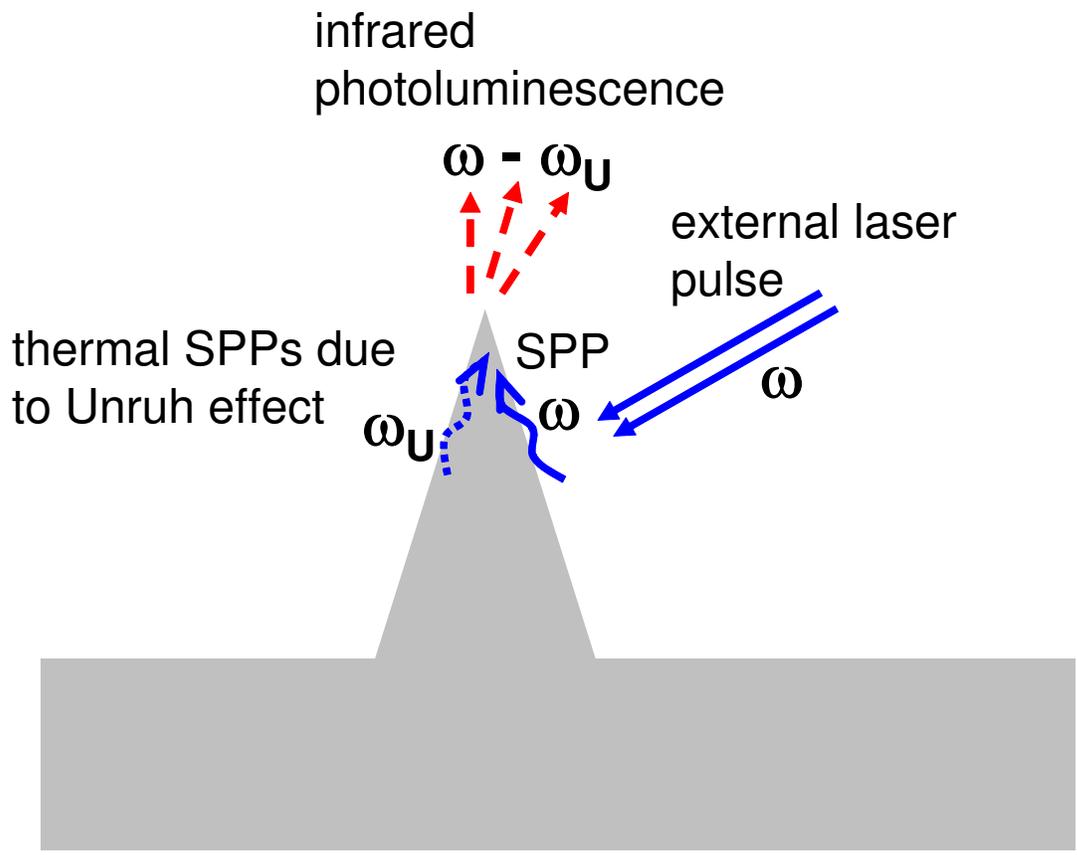
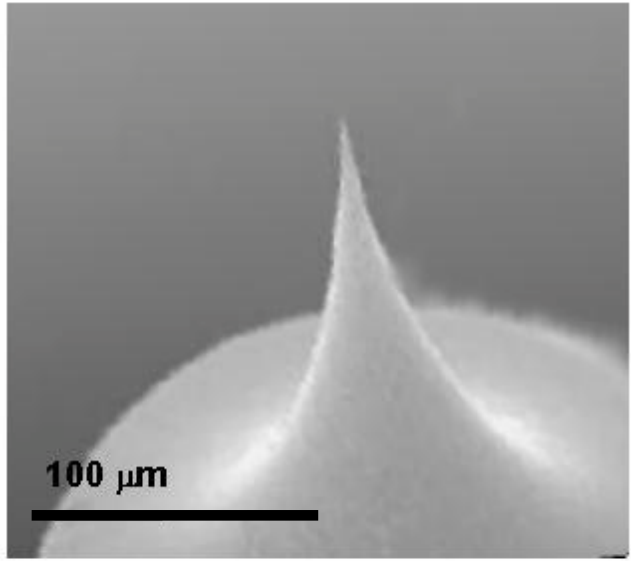
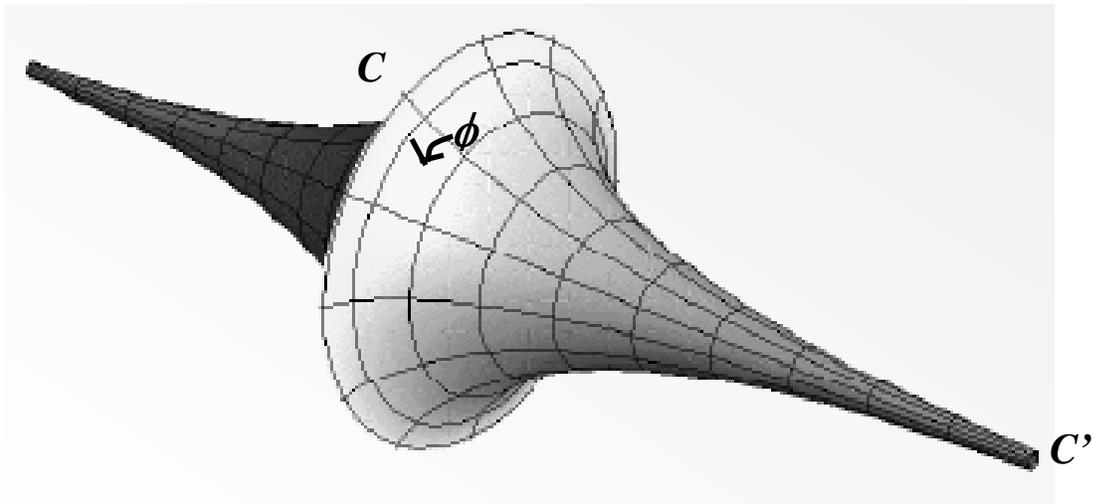
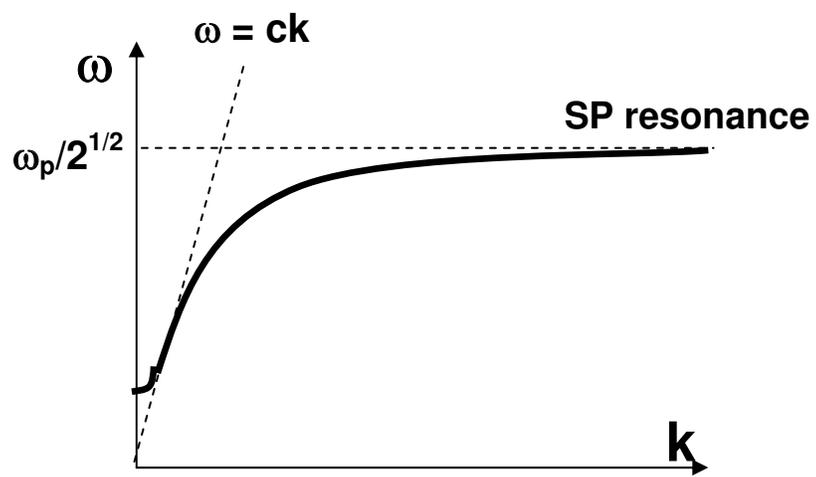
\end{document}